\documentstyle[prb,aps,epsf,multicol]{revtex}

\def\beginwideequ{
 \end{multicols}
 \vspace{-0.5truecm}
 \widetext
 \noindent
 \rule{8.7truecm}{0.1truemm}\rule{0.1 truemm}{3 truemm}\hspace*{\fill}
}
\def\endwideequ{
 \vspace{-0.2truecm}
 \hspace*{\fill}\rule{0.1 truemm}{3 truemm}\rule[3truemm]{8.7truecm}{0.1truemm}
 \vspace{-0.2truecm}
 \begin{multicols}{2}
 \narrowtext\noindent
}

\begin{document}

\tighten
\title{  Dynamical correlations in one-dimensional charge-transfer insulators}

\author{
Karlo Penc$^{1,2}$ and Walter Stephan$^{1,3}$}

\address{
$^1$Max-Planck-Institut f\"ur Physik komplexer Systeme,
N\"othnitzer Str. 38, D-01187 Dresden, Germany \\
$^2$ Research Institute for Solid State Physics and Optics,
   H-1525 Budapest, P.O.B. 49, Hungary \\
$^3$Department of Physics, Bishop's University,
Lennoxville, Qu\'ebec, Canada J1M 1Z7
 }

\date{\today}
\maketitle

\begin{abstract}
The single-particle spectral function and the density response of a two band 
Emery model 
for CuO chains is calculated for large on-site Cu repulsion $U$ and large 
on-site energy difference $\Delta$. For $U \gg U-\Delta \gg t$
the eigenfunctions are products of charge and spin parts, which allows
analytical calculation of spectral functions in that limit. 
For other parameters numerical diagonalization is used.
The low energy hole carriers are shown to be the one-dimensional analogs 
of the Zhang-Rice singlets.  The validity of the one-band model is 
discussed. The results are relevant to the interpretation of 
photoemission and EELS experiments on 
${\rm SrCuO}_2$ and ${\rm Sr}_2{\rm CuO}_3$ .
\end{abstract}


\widetext
\begin{multicols}{2}

\narrowtext

\section{Introduction}

 One of the most intriguing phenomena in one-dimensional electron systems
is the so called spin-charge separation: the low energy excitations are 
decoupled collective modes of charge and spin character, which may have 
different velocities, and are referred to as holons and spinons, 
respectively. As a consequence, the spin and charge of an added electron 
will be spatially separated after some time and there are no
Fermi-liquid like quasiparticles. 
While the decoupling exists also in the weak coupling limit\cite{solyom}, 
it is perhaps best understood for the strong coupling limit 
of the Hubbard model,
where the Bethe ansatz solution tells us that the wave functions are factorized
into a part describing free spinless fermions representing the charges and a 
part representing the spins\cite{bogyo}. This allowed the calculation of the 
dynamical spectral functions of the Hubbard model at  
\cite{sorella} and away \cite{penc} from half filling with excellent 
resolution.  These calculations provided
an explanation of the origin of the different features in the spectral
function.

The most direct test of the theory is to look at the photoemission spectra
of highly anisotropic materials. 
The nearly ideally one-dimensional CuO chains\cite{SrCuO} in the 
charge transfer insulators ${\rm SrCuO}_2$ and ${\rm Sr}_2{\rm CuO}_3$ 
are perfect candidates, given that the typical energy scale for spin and 
charge excitations is large compared to the experimental resolution, making the
observation of the low energy spectrum possible.  
The absence of bands would indicate 
that we are not dealing with the usual quasiparticles of Fermi liquid theory. 
On the other hand, there are very clear theoretical predictions for the 
photoemission spectrum of a system where spin-charge separation
exists, and indeed, recent photoemission experiments on 
${\rm SrCuO}_2$ \cite{Kim,Kim2,ttakahashi} and 
photoemission\cite{fujisawa,fujisawa2}  and electron-energy-loss\cite{neudert} 
experiments on 
${\rm Sr}_2{\rm CuO}_3$ seem to indicate that the dynamics at low energies 
can be understood within an effective $t$-$J$ or Hubbard model. 

 In comparing the measured spectra with the theoretical ones, we face the 
following difficulties: i) the actual material is a charge transfer 
insulator, while the Hubbard/$t$-$J$ model is a Mott insulator. Therefore one 
is lead to question how much of the spectra can be attributed
to generic  features where the details of the model are not important;
ii) For the CuO$_2$ plane, the $t$-$J$ model is derived to describe the 
dynamics of complex objects - the Zhang-Rice singlets\cite{ZRsinglet}.
 In the CuO$_3$ chains
the O ions are not all identical, and the original picture of Zhang and Rice
has to be refined; iii) On the theoretical side, apart from numerical 
calculations of the dynamical correlations which are difficult to interpret,
 exact and/or analytical results are very rare concerning the spectral 
function \cite{sorella,weakcoupling,kato,eder} and the optical conductivity
(e.g. for the Hubbard model see Ref.~\onlinecite{optics}).  

\begin{figure}
\epsfxsize=8 truecm
\centerline{\epsffile{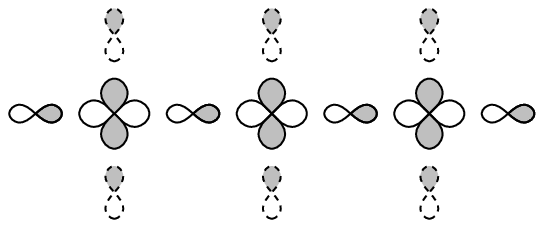}}
\caption{
The CuO$_3$ chain with 3d orbitals of Cu and 2p orbitals of O. 
The side oxygens (dashed lines) are omitted in the two-band model.
}
\label{fig:CuO}
\end{figure}

 To answer these questions, we will consider the 1D model involving the Cu 
ions and the oxygens between them (see Fig.~\ref{fig:CuO}), the so called 
two-band Emery model\cite{emery,vermeulen},  
as the simplest extension of the Hubbard model. 
In this paper we show that the charge-spin factorized wave function is an 
exact eigenfunction 
of the Emery model in the strong coupling limit, however the spinless 
fermions (`charges') represent complex objects which are the one 
dimensional analogs of the Zhang-Rice singlets. Furthermore, we  
demonstrate that the Emery model can
naturally explain the reduction of spectral weight for small momenta 
seen experimentally\cite{Kim,Kim2,ttakahashi,fujisawa,fujisawa2} and 
also describes some higher energy features of the photoemission 
spectra.

  Since the formation of Zhang-Rice singlets is an essentially strong
coupling phenomenon, the approach we present in the paper
is the most suitable method to apply. Weak coupling approaches 
\cite{weakcoupling} are inappropriate to capture the additional physics due
to the presence of additional bands.

  The outline of the paper is as follows: In Sec. II we introduce the Emery
  model and the canonical transformation leading to the strong coupling
  effective model. The spectral functions within this effective model are
  calculated in Sec. III, while in Sec. IV the density-density correlations
  relevant to the EELS experiments are discussed.

 \section{The Emery model}

\subsection{Definition of the model}

The Emery model is given by the Hamiltonian 
${\cal H} = {\cal T} + {\cal U} + {\cal V}$. 
For the kinetic part we take the usual tight-binding form,
\begin{equation}
 {\cal T} = -t \sum_{i,\delta,\sigma} 
    (d^{\dagger}_{i,\sigma} p^{\phantom{\dagger}}_{i+\delta,\sigma} 
     + {\rm H.c.} ), 
\end{equation}
where $d^{\dagger}_{i,\sigma}$ and 
$p^{\dagger}_{i+\delta,\sigma}$ denote the 
hole creation operators on copper $d$ and oxygen $p$ 
orbitals at sites $i$ and $i+\delta$, respectively. 
The Cu-Cu distance is taken to be unity, $i$ are integers and $\delta=\pm 1/2$.
 The phase factors in the hybridization coming from the symmetry of the Cu and
 O orbitals are absorbed in the definition of the $d$ and $p$ operators as 
$d_j=(-1)^j d_{{\rm phys},j}$, and $p_{j+1/2}=(-1)^j p_{{\rm phys},{j+1/2}}$,
 where the 
subscript `phys' denotes the operators respecting the phase factors in 
the hybridization. The inclusion of the phase factors causes a shift 
of $\pi$ in the momentum of the $O$ hole and will be explicitly
mentioned when necessary. 
In the potential part we include the on-site energy difference 
$\Delta = \varepsilon_p - \varepsilon_d$ and the on-site Coulomb repulsion 
$U$ of the Cu $3d$-orbitals:
\begin{equation}
 {\cal U} =  \frac{\Delta}{2} 
     \sum_{i} \left(n^p_{i+1/2} - n^d_{i} \right)
     + U \sum_{i} n^d_{i,\uparrow} n^d_{i,\downarrow} ,
\end{equation}
where  $n^a_{i,\sigma} =
a^{\dagger}_{i,\sigma}  a^{\phantom{\dagger}}_{i,\sigma}$ ($a=d,p$), and 
$n^a_i = \sum_\sigma n^a_{i,\sigma}$, furthermore the nearest neighbor Cu-O 
repulsion 
\begin{equation}
 {\cal V} =  V \sum_{i,\delta} n^d_{i} n^p_{i+\delta},
\end{equation}
which may lead to exciton formation\cite{vermeulen}. We choose 
$U>\Delta$ in order to have a charge transfer insulator  with one hole per 
unit cell\cite{sawatzki}. 

Note that if one begins with a model which also includes 
oxygen orbitals on the side of the chain, as a preliminary step
one may build bonding, and anti- and non-bonding combinations of these.
Therefore, if one understands our single oxygen orbital per cell to correspond
to the bonding combination, our results may also be seen to represent
a good approximation to part of the spectrum of the more complete
four band model\cite{rosner,Drechsler}.

 \subsection{Effective model in the strong coupling limit}
As mentioned earlier, direct numerical methods, such as exact diagonalization, 
work only for rather small system sizes. 
However, in the strong coupling limit ($U,\Delta\gg t,V$) it is  possible
to do controlled calculations both analytically and numerically.
As a first step, we derive an effective strong coupling Hamiltonian.
In the extreme case when $t=V=0$, the Hamiltonian is block-diagonal in the 
subspace of states with a given eigenvalue of ${\cal U}$.
The hybridization in ${\cal T}$ may be treated perturbatively using a 
canonical transformation \cite{harris}, leading to an effective Hamiltonian 
${\cal H}_{\rm eff}$ acting within one subspace.
A detailed and
systematic explanation for the case of the Hubbard model is given in 
Ref.~\onlinecite{oles}. We denote by tilde (e.g. $\tilde p$) the operators 
acting in the 
subspace of ${\cal H}_{\rm eff}$, and the physical operators ${\cal O}$ 
are then obtained from
\begin{equation} 
{\cal O} = e^{\cal S} \tilde {\cal O}  e^{-{\cal S}} = 
 \tilde {\cal O} + [ {\cal S},\tilde {\cal O} ] + \cdots,
\end{equation}
 where 
$\tilde {\cal O} \equiv {\cal O}(p\rightarrow \tilde p,d\rightarrow \tilde
d,\dots)$ and 
\begin{equation}
  {\cal S}= \frac{1}{\Delta}
       \left(\tilde {\cal T}^{\phantom{\dagger}}_\Delta 
         - \tilde {\cal T}^{\dagger}_{\Delta}\right) + 
       \frac{1}{U-\Delta}
       (\tilde {\cal T}^{\phantom{\dagger}}_{U-\Delta} 
         - \tilde {\cal T}^{\dagger}_{U-\Delta}) +O(t^2)
\end{equation}
is the generator of the canonical transformation with 
\begin{eqnarray}
 \tilde {\cal T}_\Delta &=& -t \sum_{i,\delta,\sigma} 
    \tilde p^{\dagger}_{i+\delta,\sigma} 
    \tilde d^{\phantom{\dagger}}_{i,\sigma} 
    (1-\tilde n^d_{i,\bar\sigma}), 
 \nonumber\\
 \tilde {\cal T}_{U-\Delta} &=& -t \sum_{i,\delta,\sigma} 
    \tilde d^{\dagger}_{i,\sigma} 
    \tilde p^{\phantom{\dagger}}_{i+\delta,\sigma} 
    \tilde n^d_{i,\bar\sigma}.
\end{eqnarray}
The subscript $nU+m\Delta$ denotes that the state acted upon is
promoted to a subspace at this energy difference. In other words 
\begin{equation}
 \bigl[ \tilde {\cal U}, {\cal O}_{nU+m\Delta} \bigr] 
 = (nU+m\Delta) {\cal O}_{nU+m\Delta},
\end{equation}
and every operator can be decomposed
as  ${\cal O} = \sum_{n,m}{\cal O}_{nU+m\Delta}$
 with $n$ integers and $m$ integers or half-odd integers. 
Then ${\cal H}_{\rm eff}$ is given by:
\begin{eqnarray}
   {\cal H}_{\rm eff} &=& \tilde {\cal U} + \tilde {\cal V} 
    + \frac{1}{\Delta} 
      \left[
        \tilde {\cal T}^{\phantom{\dagger}}_\Delta,
        \tilde {\cal T}^{\dagger}_{\Delta}
      \right] 
    + \frac{1}{U-\Delta} 
      \left[
        \tilde {\cal T}^{\phantom{\dagger}}_{U-\Delta},
        \tilde {\cal T}^{\dagger}_{U-\Delta}
      \right]\nonumber\\ 
   &&+O(t^2).
\end{eqnarray}
Separating the different processes, 
${\cal H}_{\rm eff} ={\cal U}' + {\cal H}_{0} + 
     {\cal H}_{1} + {\cal H}_{2}$, and introducing the effective hopping 
amplitudes $t_S = t^2/(U-\Delta)$ and $t_T = t^2/\Delta$, 
furthermore $U' = U+ 4 t_T + 4 t_S$ and 
$ \Delta' = \Delta + 4 t_T$, we get 
(see also Refs. \onlinecite{canpert,reiter}):
\end{multicols}
 \widetext
 \noindent
\begin{eqnarray}
 {\cal U}'  &=&
    \frac{\Delta'}{2} 
    \sum_{i} (\tilde n^p_{i+1/2} \!-\! \tilde n^d_{i})
    + U' 
    \sum_{i} \tilde n^d_{i,\uparrow} \tilde n^d_{i,\downarrow} 
    + V \sum_{i,\delta}  \tilde n^d_{i} \tilde n^p_{i+\delta},
 \nonumber\\
{\cal H}_{0}  &=&    - t_T 
      \sum_{i,\sigma,\delta}
       (1 \!-\! \tilde n^d_{i+2\delta,\bar\sigma}) 
        \tilde d^{\dagger}_{i+2\delta,\sigma} 
        \tilde d^{\phantom{\dagger}}_{i,\sigma} 
        (1 \!-\! \tilde n^d_{i,\bar\sigma}) ,
  \nonumber\\
{\cal H}_{1}  &=&
      \left( t_S + t_T \right)
      \sum_{i,\delta,\delta',\sigma}
        \left(
          \tilde p^{\dagger}_{i+\delta,\sigma} 
          \tilde d^{\dagger}_{i,\bar\sigma} 
          \tilde d^{\phantom{\dagger}}_{i,\sigma} 
          \tilde p^{\phantom{\dagger}}_{i+\delta',\bar\sigma}
        - \tilde p^{\dagger}_{i+\delta,\sigma} 
          \tilde n^d_{i,\bar\sigma} 
          \tilde p^{\phantom{\dagger}}_{i+\delta',\sigma} 
        \right) 
      + t_T 
      \sum_{i,\sigma,\delta}
        \tilde p^{\dagger}_{i+\delta,\sigma} 
        \tilde p^{\phantom{\dagger}}_{i-\delta,\sigma} ,
\nonumber\\
{\cal H}_{2}  &=&  t_S
      \sum_{i,\sigma,\delta}
        \tilde n^d_{i+2\delta,\bar\sigma} 
        \tilde d^{\dagger}_{i+2\delta,\sigma} 
        \tilde d^{\phantom{\dagger}}_{i,\sigma}
        \tilde n^d_{i,\bar\sigma} .
\end{eqnarray}
\endwideequ
${\cal H}_0$ and ${\cal H}_2$ describe the motion of the empty 
and doubly occupied site, respectively, while ${\cal H}_1$ is 
responsible for the dynamics of the hole on oxygen.
 By using the Heisenberg model ground state $|{\rm GS}\rangle$
for the insulating case, the main effect of the (fourth order) AF 
interaction\cite{hirsch} between Cu spins is accounted for.

 \section{spectral functions}
 
Now let us turn to the spectral functions. The 
photoemission spectrum is proportional to the 
single particle spectral function, defined by
\begin{eqnarray}
  B(k,\omega) &=& \sum_{f,\sigma} 
        |\langle f | p^\dagger_{k,\sigma} |{\rm GS} \rangle|^2
  \delta(\omega+E_f-E_{\rm GS}) \nonumber\\
 && + \sum_{f,\sigma} 
        |\langle f | d^\dagger_{k,\sigma} |{\rm GS} \rangle|^2
  \delta(\omega+E_f-E_{\rm GS}) ,
 \label{eq:bkw}
\end{eqnarray}
assuming that the cross sections of the Cu and O electron removal are equal.
The sum is over final states $|f\rangle$ with a hole added, and a 
similar definition holds for the inverse photoemission spectra, where a
 hole is removed. First, we will present the analytical and numerical
 calculation of the spectral function for the effective model, and then 
we will compare our results to the spectral function of $t$-$J$ model and the
photoemission spectra of SrCuO$_2$ and Sr$_2$CuO$_3$.

\subsection{Calculation of spectral functions}

\begin{figure}
\epsfxsize=8 truecm
\centerline{\epsffile{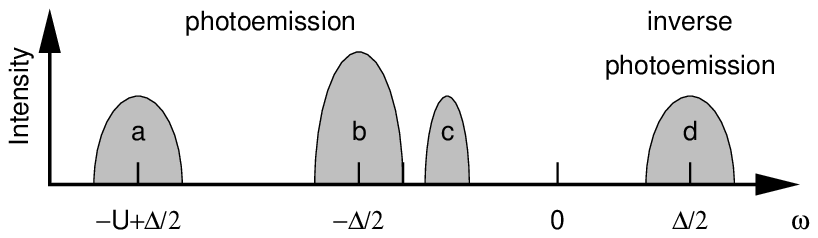}}
\caption{
Schematic distribution of the weights in the strong coupling limit for a
charge transfer insulator.
}
\label{fig:schema}
\end{figure}

Since the nearest neighbor Coulomb repulsion leads only to a uniform
shift of the final state energies in the spectral functions, it
will be neglected in this section.
The strong coupling behavior of the photoemission spectra is schematically
shown in Fig.~\ref{fig:schema}: the hole can go either to the Cu site (peak
`a') or to the O site (peaks `b' and `c').
The creation operator 
$c^\dagger_{k,\sigma} = d^\dagger_{k,\sigma},p^\dagger_{k,\sigma}$ 
entering  the calculation of the spectral function 
Eq.~(\ref{eq:bkw}) can be decomposed in leading order as 
$c^{\dagger}_{k,\sigma} = c^{\dagger}_{k,\sigma;-\Delta/2} 
  + c^{\dagger}_{k,\sigma;\Delta/2} + c^{\dagger}_{k,\sigma;U-\Delta/2}$ 
which represent a hybridized
mixture of Cu and O atomic states. For example, including $O(t)$ corrections
\begin{eqnarray}
  p^{\dagger}_{i+\delta,\sigma;\Delta/2} &=& 
   \tilde p^{\dagger}_{i+\delta,\sigma} +O(t^2),\\
  d^{\dagger}_{i,\sigma;\Delta/2} &=& 
   - \frac{t}{\Delta} 
     \sum_{\delta} 
    [  (1-\tilde n^d_{i,\bar\sigma}) 
       \tilde p^{\dagger}_{i+\delta,\sigma} 
       - \tilde d^{\dagger}_{i,\sigma}
         \tilde p^{\dagger}_{i+\delta,\bar\sigma}
         \tilde d^{\phantom{\dagger}}_{i,\bar\sigma}
    ] \nonumber\\
   &&+ \frac{t}{U-\Delta} 
     \sum_{\delta} 
    [  \tilde n^d_{i,\bar\sigma}
       \tilde p^{\dagger}_{i+\delta,\sigma} 
       + \tilde d^{\dagger}_{i,\sigma}
         \tilde p^{\dagger}_{i+\delta,\bar\sigma}
         \tilde d^{\phantom{\dagger}}_{i,\bar\sigma}
    ] \nonumber\\
   && + O(t^2).
  \label{eq:ptilde2p}
\end{eqnarray}
The final states in peaks `b' and `c' are obtained
by applying $c^{\dagger}_{k,\sigma;\Delta/2}$ to the ground state, and 
the sum rule 
of the `b' + `c' peaks in Fig.~\ref{fig:schema} is
\begin{eqnarray}
  && \langle {\rm GS} | 
     p^{\phantom{\dagger}}_{k,\sigma;\Delta/2}
     p^{\dagger}_{k,\sigma;\Delta/2} | {\rm GS} 
   \rangle + \nonumber\\
  && \langle {\rm GS} | 
     d^{\phantom{\dagger}}_{k,\sigma;\Delta/2}
     d^{\dagger}_{k,\sigma;\Delta/2} | {\rm GS} 
   \rangle = 1+ O(t^2).
\label{eq:sumrule}
\end{eqnarray}
Similarly, the weight in peak `a' and `d' is  $1/2+O(t^2)$.
For simplicity, we do not include the corrections in Eq.~(\ref{eq:ptilde2p}) 
when we calculate the spectral functions.

The dynamics in the peak `a' 
is governed by ${\cal H}_2$: the 
extra hole on Cu, created by 
$\tilde d^{\dagger}_{k,\sigma}$ in 
Eq.~(\ref{eq:bkw}), hops to neighboring Cu sites with
amplitude $t_S$, leaving the spin sequence unchanged:
\begin{eqnarray} 
 {\cal H}_2 |\cdots \sigma_{j-1} 2 \sigma_{j} \cdots \rangle &=& -t_S 
  |\cdots 2 \sigma_{j-1} \sigma_{j} \cdots \rangle \nonumber\\
 && - t_S|\cdots \sigma_{j-1} \sigma_{j} 2 \cdots \rangle . \label{eq:h22}
\end{eqnarray}
Here `2' in the wave function denotes the position of the extra hole 
(site occupied by two holes), and $\sigma_j$ the spins of the 
singly occupied Cu sites with $j=1,\dots,L-1$, as there are $L-1$ spins 
remaining.  
The situation is identical to the case of $U/t \to +\infty$ Hubbard model, 
where we know that the wave function of a state with momentum $k$ 
factorizes into charge and spin parts\cite{bogyo,sorella,penc}:
\begin{equation}
 | f_Q(k) \rangle = \frac{1}{\sqrt{L}}
   \sum_{j=0}^{L-1} e^{i(k-Q)j} | \psi_j \rangle 
   \otimes | \chi_{L-1}(Q,n_Q) \rangle ,
 \label{eq:facwf}
\end{equation}
where $| \psi_j \rangle$ describes free spinless fermions on $L$ lattice 
points with an empty site $j$, which in our case is the site with the extra hole. 
$| \chi_{L-1}(Q,n_Q) \rangle$ is the squeezed 
spin wave function of $L-1$ spins with momentum $Q=2\pi J/(L-1)$, $J$ integer,
 and other quantum numbers $n_Q$.
 The energy of the state is 
\begin{equation}
 \varepsilon_{Q}(k) = U'-\Delta'/2  - 2t_S \cos(k-Q).
\end{equation}
Now that we have both the energy and the wave function of the final 
state, we are ready to calculate the spectral function as presented by 
Sorella and Parola for the large-$U$ Hubbard 
model\cite{sorella}. As a first step, we write the ground 
state also in a product form
\begin{equation} 
 |{\rm GS} \rangle = | \psi_{\rm GS} \rangle \otimes | \chi_{\rm GS} \rangle ,
\end{equation}
where the $| \chi_{\rm GS} \rangle$ is the Heisenberg ground-state wave function 
and $| \psi_{\rm GS} \rangle$ is the fully filled Fermi see of 
spinless fermions (charges). It is convenient to choose systems with 2,6,10 
etc. sites, where the momentum of the $|{\rm GS} \rangle$ is zero. 
In the matrix element of Eq.~(\ref{eq:bkw}) it suffices to keep the 
momentum dependence in the final wave function only,
\begin{eqnarray}
  \bigl| \langle f_Q(k) | \tilde d^{\dagger}_{k,\sigma} | {\rm GS } \rangle 
  \bigr|^2 &=& 
  L \bigl| \langle f_Q(k) | \tilde d^{\dagger}_{j=0,\sigma} | {\rm GS } \rangle 
  \bigr|^2 \nonumber\\
  &=& \bigl|
       \langle \chi(Q,n_Q) | Z_{0,\sigma} | \chi_{\rm GS } \rangle 
  \bigr|^2,
\end{eqnarray}
where we have substituted the factorized form Eq.~(\ref{eq:facwf}) and
used the fact that the overlap in the charge part is 1. 
Only the spin part is nontrivial:
the operator $Z_{j,\sigma}$ removes a spin
$\sigma$ at site $j$, reducing the spin sequence to length $L-1$. Introducing 
\begin{equation} 
  D(Q) = \sum_{n_Q} \bigl| \langle \chi(Q,n_Q) | Z_{0,\sigma} 
  |\chi_{\rm GS} \rangle \bigr|^2
\end{equation} 
for the spectral function we get:
\begin{equation} 
   B(k,\omega) = 
   \sum_{Q} D(Q) \delta(\omega+\varepsilon_{Q}(k)).
 \label{eq:bkwuhb}
\end{equation}
$D(Q)$ is essentially the ``occupation number'' of the spinons, and has a 
singularity at the spinon fermi momenta $Q=\pm \pi/2$. It can be
approximated as $(L-1)D(Q) \approx -0.5+2.98/\sqrt{\pi^2-4Q^2}$ 
for $-\pi/2<Q<\pi/2$ and zero otherwise \cite{sorella,penc,strong}. 
 We therefore find that the spectral function in the upper Hubbard band 
(peak `a')
is identical to that of the  large-$U$ Hubbard  or small $J$  $t$-$J$ model.
Note also that the inverse photoemission spectrum will have a similar form with 
bandwidth $4 t_T$ (peak `d' in Fig.~\ref{fig:schema}).

\begin{figure}
\epsfxsize=6 truecm
\centerline{\epsffile{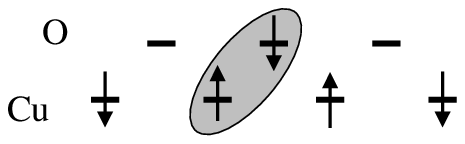}}
\caption{
The state $|\downarrow\roarrow S \uparrow\downarrow\rangle$. The Cu and O 
spins in the shaded region form a 
singlet, which we denoted by $\roarrow S$.
}
\label{fig:singlet}
\end{figure}

 Let us now proceed to peaks `b' and `c' in Fig.~\ref{fig:schema}, which 
can be associated with the hole on oxygens. The hole added to an O site 
with $\tilde p^{\dagger}_{k,\sigma}$ 
can form a singlet or triplet with a neighboring Cu spin. 
We will denote by $|\loarrow S \rangle$ ($|\roarrow S \rangle$)
states where the O hole forms a singlet with the Cu spin on its
right (left), as seen in Fig.~\ref{fig:singlet}. For example,
\begin{eqnarray}
  |\downarrow \roarrow S \uparrow \downarrow \rangle &=&\frac{1}{\sqrt{2}}
   \tilde d^\dagger_{1\downarrow} 
   \bigl( \tilde d^\dagger_{2\uparrow} \tilde p^\dagger_{5/2\downarrow} -
          \tilde d^\dagger_{2\downarrow} \tilde p^\dagger_{5/2\uparrow} \bigr)
   \tilde d^\dagger_{3\uparrow} \tilde d^\dagger_{4\downarrow}  | 0 \rangle ,
\nonumber\\
  |\downarrow \loarrow S \uparrow \downarrow \rangle &=&\frac{1}{\sqrt{2}}
    \tilde d^\dagger_{1\downarrow} 
   \bigl( \tilde p^\dagger_{3/2\downarrow}\tilde  d^\dagger_{2\uparrow} -
          \tilde p^\dagger_{3/2\uparrow} \tilde d^\dagger_{2\downarrow} \bigr)
  \tilde  d^\dagger_{3\uparrow} \tilde d^\dagger_{4\downarrow}  | 0 \rangle .
  \nonumber
\end{eqnarray} 

As we will see below,
a suitable combination of these states will give us the 
Zhang-Rice singlets\cite{ZRsinglet}, 
in terms of which the lowest energy excitations may be described
by a one-band model. In the present calculation we are also 
including the triplets and high-energy singlets in order to describe 
higher energy excitations. 
Note that this basis is not orthogonal:
\begin{eqnarray}
  \langle \cdots \roarrow S \sigma \cdots|
     \cdots \roarrow S  \sigma \cdots \rangle &=& 1, \nonumber\\
  \langle\cdots \roarrow S \sigma \cdots|
 \cdots\sigma\loarrow S \cdots\rangle &=& 1/2.
\end{eqnarray} 
In general, the resulting spectrum is complicated and the singlets and 
triplets mix with one other,  except for the 
particular case of $t_S$ finite and $t_T=0$ \cite{reiter,FCZhang}.  
Due to the very special form of  ${ \cal H}_1$ for $t_T=0$
 \begin{equation}
  { \cal H}_1  =
     - 2 t_S 
      \sum_{i,\delta,\delta',\alpha,\alpha'}
        \left( \frac{\delta_{\alpha\alpha'}}{4}  \tilde n_i^d
            -  \frac{\bbox{\tau}_{\alpha\alpha'}}{2} {\bf \tilde S}_i^d
        \right) 
          \tilde p^{\dagger}_{i+\delta,\alpha} 
          \tilde p^{\phantom{\dagger}}_{i+\delta',\alpha'} ,
\end{equation}
where we can identify the projector onto spin singlets 
(here ${\bf \tilde S}_i^d = \sum_{\sigma,\sigma'} 
\tilde d^{\dagger}_{i,\sigma} 
 \bbox{\tau}_{\sigma\sigma'}  \tilde d^{\phantom{\dagger}}_{i,\sigma'}/2 $ 
and  $ \bbox{\tau}$ is the vector of Pauli matrices), 
the matrix elements of ${\cal H}_{\rm eff}$ leading to propagation of the
singlets are:
\begin{eqnarray}
{\cal H}_1 |\cdots\roarrow S \sigma \cdots \rangle &=& 
     t_S | \cdots\sigma (\roarrow S-\loarrow S) \cdots\rangle 
 \nonumber\\&&
   - 2 t_S  |\cdots (\roarrow S-\loarrow S)  \sigma \cdots\rangle,
\nonumber\\
{\cal H}_1 |\cdots\sigma \loarrow S \cdots \rangle &=& 
   -t_S | \cdots (\roarrow S-\loarrow S)\sigma \cdots\rangle
 \nonumber\\&&
   + 2 t_S |\cdots \sigma(\roarrow S-\loarrow S)  \cdots\rangle . 
 \nonumber
\end{eqnarray}
The combination 
$|\roarrow S\rangle - |\loarrow S\rangle$ moves through the lattice like
the site `2' in Eq.~(\ref{eq:h22})
\begin{eqnarray}
{\cal H}_1 |\cdots \sigma_{j-1} (\roarrow S - \loarrow S) \sigma_j \cdots 
\rangle &=& t_S 
  \Bigl( 
        | \cdots\sigma_{j-1}\sigma_j (\roarrow S-\loarrow S) \cdots\rangle
  \nonumber\\ 
   &&  - 4  |\cdots \sigma_{j-1}(\roarrow S-\loarrow S) \sigma_j \cdots\rangle 
       \nonumber\\ 
   && + |\cdots (\roarrow S-\loarrow S)\sigma_{j-1}\sigma_j \cdots\rangle
  \Bigr), \nonumber\\ 
   &&
\end{eqnarray}
 leaving the spin sequence unchanged.  In this case $| \psi_j \rangle$ in 
Eq.~(\ref{eq:facwf}) will denote this particular combination at site $j$.
The energy of the state is 
\begin{equation}
 \varepsilon^S_{Q}(k) = \Delta/2 
 - 4 t_S + 2 t_S \cos(k-Q) . 
\end{equation}
These singlets leads to the formation of the `c' peak in 
Fig.~\ref{fig:schema}. 

Next, we need to calculate the matrix elements. Using the identity 
\begin{equation}
{\bf |} \langle f | \tilde p^\dagger_{k,\sigma} | {\rm GS} \rangle {\bf |}^2 
  = L {\bf |} \langle{\rm GS}| \tilde p_{1/2,\sigma} | f(k) \rangle {\bf |}^2,
\end{equation}
where $p_{1/2,\sigma}$ removes the hole at site $i=1/2$,
the $k$ dependence is now in the final state only. So, for the matrix 
element we get:
\beginwideequ
\begin{eqnarray}
 L {\bf |}\langle{\rm GS}| \tilde p_{1/2,\downarrow} | f(k) \rangle {\bf |}^2 
  &=& 
    \Bigl| \sum_j e^{i(k-Q)j}
        \bigl(\langle \chi_{\rm GS} | \otimes \langle \psi_{\rm GS} |\bigr)
        \tilde p_{1/2,\downarrow}
      \bigl(| \psi_j \rangle \otimes |\chi(Q,n_Q) \rangle \bigr)\Bigr|^2
 \nonumber\\
   &=& 
    \frac{1}{2} \bigl|\langle \chi_{\rm GS}| 
  \bigl( Z^{\dagger}_{0,\uparrow} - e^{i(k-Q)}Z^{\dagger}_{1,\uparrow}\bigr)
      |\chi(Q,n_Q) \rangle \bigr|^2
 \nonumber\\
   &=&   
    (1+\cos k) \bigl| 
      \langle \chi_{\rm GS}| Z^{\dagger}_{0,\uparrow}|\chi(Q,n_Q) \rangle 
    \bigr|^2,
 \end{eqnarray}
\endwideequ
where we have used that $\langle \chi_{\rm GS}| Z^{\dagger}_{1,\uparrow}
|\chi(Q) \rangle  =- e^{iQ} \langle \chi_{\rm GS}|
Z^{\dagger}_{0,\uparrow}|\chi(Q) \rangle$. Dividing the matrix element by 
the norm of the final state 
\begin{equation} 
  \langle f_Q(k) | f_Q(k) \rangle = 2-\cos (k-Q) = 
  \frac{\Delta - 2\varepsilon^S_{Q}(k)}{4t_S},
\end{equation} 
and summing over final states with definite $Q$, we can write the spectral 
function as
\begin{equation}
   B_S(k,\omega) = 
    \frac{4t_S(1+\cos k)}{\Delta + 2\omega}
   \sum_{Q} D(Q) \delta(\omega+\varepsilon^S_{Q}(k)).
 \label{eq:aoksingl}
\end{equation}

Clearly, even introducing form factors in the one-band model 
[which is identical in form to Eq.~(\ref{eq:bkwuhb})], the
$\omega$-dependent prefactor of the spectral
distribution above cannot be obtained. 
The local ($k$-averaged) spectral function for the singlets is
\begin{equation} 
   B_S(\omega) = 
     \frac{1}{\pi}
     \frac{4 t_S + (2 \ln 2 \!-\!1)(-2 \omega \!-\! \Delta \! + \! 8t_S)}
          {(\Delta \! + \! 2\omega)
           \sqrt{(\Delta \! - \!4t_S \! + \! 2\omega)
           (-2\omega \! - \! \Delta \! + \! 12t_S)}},
\end{equation}
with weight 0.32. The rest of the weight (0.68) is  
at higher energies $\omega=-\Delta/2$, where we find nondispersing 
solutions, made of a particular combination of singlets
$| \roarrow S \sigma \sigma' \rangle 
 - 2 | \sigma \roarrow S \sigma' \rangle
 - 2 | \sigma \loarrow S \sigma' \rangle
   + | \sigma \sigma' \loarrow S \rangle$, as well as the  triplets, contributing 
with a delta peak to the spectral function to form the peak `b' in 
Fig.~\ref{fig:schema}.

  The only requirement for the procedure outlined above to work is that during 
the motion of the hole the spin sequence is unchanged. This immediately 
requires $t_T=0$ in ${\cal H}_1$.

  The influence of finite $t_T$ is shown in Fig.~\ref{fig:aok}. The lower 
`singlet' band increases its width, 
while the overall shape of 
$B_S(k,\omega)$ does not change significantly. On the other hand,
peak `b' now extends from $\Delta/2$ to $\Delta/2+4t_T$ and a 
sharp dispersion dominates the spectrum.
Only a slight weight 
transfer from the `singlet' to the `triplet' band can be observed, 
e.g. at $k=0$ ($q=\pi$) the weight in the 
`singlet' band is reduced from 0.65 to 0.43 as we 
increase $t_T$ from 0 to $t_S$, 
while the total weight, given by Eq.~(\ref{eq:sumrule})
is unchanged in leading order.

  \subsection{Comparison with the t-J model and photoemission experiments}
\begin{figure}
\epsfxsize=15 truecm
\centerline{\epsffile{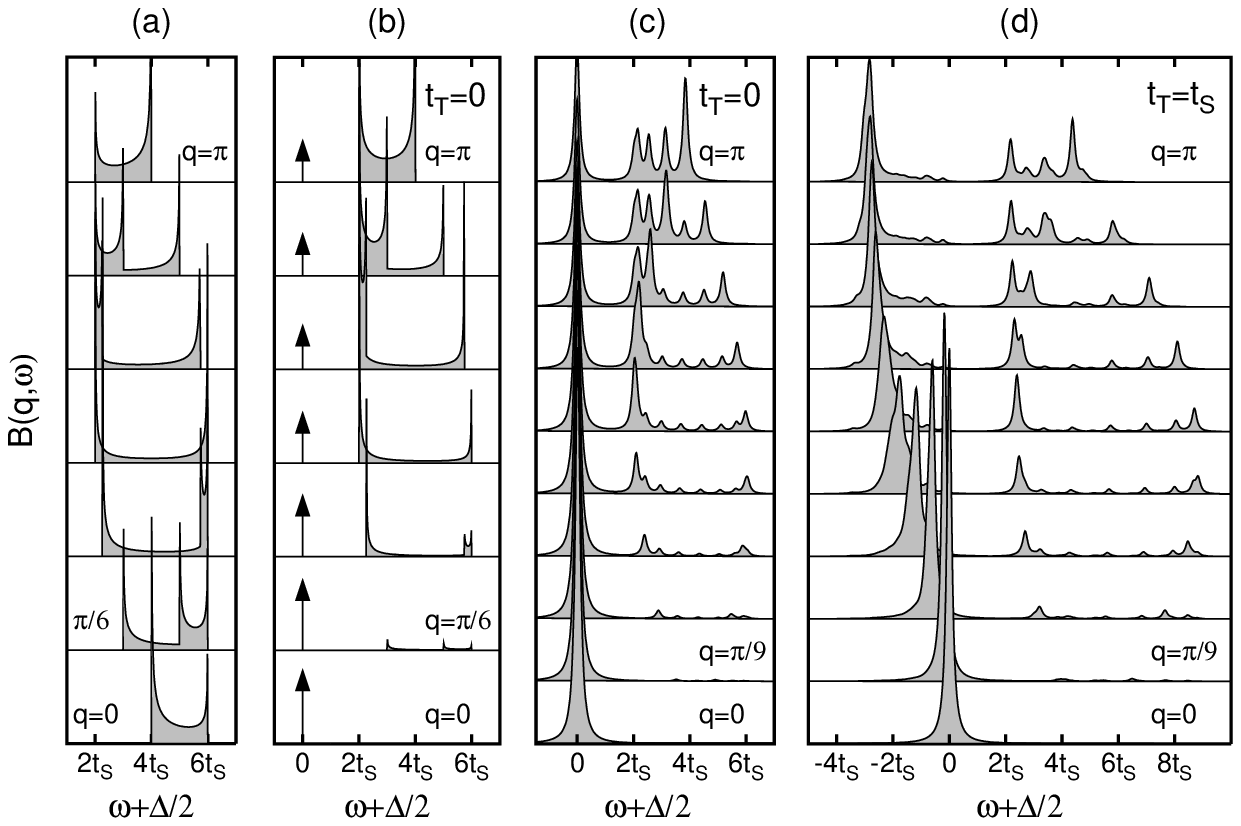}}
\caption{
 The analytical result for photoemission spectra of (a) $t-J$ model in the 
$J/t \to 0$ limit [Eq.~({\protect\ref{eq:aoksingl}})  without the $k$ and 
$\omega$ dependent prefactors] and (b) $t_T=0$ effective model 
[Eq.~({\protect\ref{eq:aoksingl}})],  compared 
to (c) a L\'anczos diagonalization of 18 site effective model for $t_T=0$  
and (d) $t_T=t_S$.
 The $\delta$ functions are plotted as Lorentzians of width $0.1$. The 
$q=k+\pi$ is the momentum when the 
relative phases of $d$ and $p$ orbitals are properly included and it should be
used when we compare with the experiments.}
  \label{fig:aok}
\end{figure}

Comparing with the $t$-$J$ model for small $J$ (Fig.~\ref{fig:aok}a), 
we can see that although (also for $t_T=t_S$) the `singlet' feature 
is similar to the $t$-$J$ model result \cite{sorella}, 
there are detailed differences in the distribution of weight, similar to 
those in Eq.~(\ref{eq:aoksingl}), as well as in the dispersion of the 
upper edge of the `singlet' continuum. We therefore see that even in 
parameter regimes where the one-band $t$-$J$ description accurately 
predicts low-energy excitation energies, the two-band model may have 
significantly different properties as far as other physical observables 
is concerned, exemplified here by the momentum and frequency dependence
of the spectral weights.
The effect of finite $J$ is to give dispersion to the
now dispersionless lower `spinon' edge in both Emery and $t$-$J$ model.

Finally, let us compare our results with the experiments.
For both ${\rm SrCuO}_2$ and ${\rm Sr}_2{\rm CuO}_3$ the low energy region
shows features found in the $t$-$J$ model, i.e. the holon and 
spinon bands dispersing with $t\approx 0.5-0.6$eV and $J\approx 0.15-0.2$eV,
respectively, which is consistent with the susceptibility\cite{SrCuO}, 
optical\cite{infra} and electron-energy loss\cite{neudert} experiments.
However, an additional interesting feature is the weight reduction as 
the zone center ($q=0$) is approached. 
In Refs.~\onlinecite{Kim2,fujisawa2} this is attributed to the 
different cross sections of Cu and O orbitals, while in our theory it 
arises quite naturally from the internal structure of the low energy
singlets. 
Concerning the higher energy features, the `triplet' feature 
is in reasonable agreement with the dispersing peak at 2 eV in Fig. 6 of 
 Ref.~\onlinecite{ttakahashi}, if one disregards the flat non-bonding 
oxygen bands not included in our model. 
These general trends don't 
depend strongly on $t_T/t_S$ and the inclusion of $t/\Delta$ and $t/(U-\Delta)$
correction in operators leads only to a small weight transfer to 
lower energies. 

\end{multicols}
\widetext

\begin{multicols}{2}
\narrowtext

\section{Density-density correlations}

  The density-density correlation function describes the dynamical 
dielectric response of the material and is accessible by measuring  e.g. 
the optical conductivity, electron-energy loss spectra (EELS), and 
inelastic X-ray scattering. Both optical conductivity \cite{maiti} 
and EELS \cite{neudert} 
have been measured on Sr$_2$CuO$_3$. The EELS spectra can be reasonably well
interpreted within a one band Hubbard model extended with nearest neighbor 
repulsion, so it is interesting to see what changes if we consider a two band 
model, like the Emery model.

Since the system is an insulator, we get finite density response only above 
the charge transfer gap at $\omega\approx\Delta$. 
In lowest order the density-density correlation function is 
given by:
\begin{equation}
 {\cal N}(k,\omega) = \sum_{f} 
   |\langle f | n_{k;\Delta} | {\rm GS} \rangle|^2 
   \delta(\omega-E_f+E_{\rm GS}),
  \label{eq:rhorho}
\end{equation}
where $n_{k;\Delta} = n^d_{k;\Delta} + n^p_{k;\Delta}$  
can be calculated 
from $n_{i;\Delta}=(1/\Delta) [\tilde {\cal T}_\Delta,\tilde n_i]$ and reads
\begin{equation}  n_{k;\Delta} = \frac{t}{\Delta}\frac{1}{\sqrt{L}} 
  \sum_{j,\delta,\sigma} e^{i k j} (e^{i\delta k}-1)
         (1-\tilde n_{j,\bar \sigma})      
         \tilde p^{\dagger}_{j+\delta,\sigma} 
         \tilde d^{\phantom{\dagger}}_{j,\sigma}. 
\end{equation}
This leads to the sum rule  
\begin{equation}
\int {\cal N}(k,\omega) d\omega  = 8 \frac{t^2}{\Delta^2} \sin^2\frac{k}{4} .
\end{equation}

 Now let us determine the dynamical density response in the special case $t_T=0$.
 The operator $n_{k;\Delta}$
moves a hole from Cu to O, and results in a two body problem, which can be 
solved using standard techniques. For the `singlet' part, the 
final  state wave function can be represented as 
\begin{equation}
 | f_S \rangle =
  \sum_{j=1}^{L-1} \Bigl(
               x_j 
              | e \sigma_1 \cdots \sigma_{j-1} 
                \loarrow S \sigma_j \cdots \rangle
              - y_j 
              | e \sigma_1 \cdots \sigma_{j-1} 
                \roarrow S \sigma_j \cdots \rangle \Bigr),
\end{equation}
with the norm
\begin{equation}  
  \langle f_S | f_S \rangle = \sum_{j=1}^{L-1} (x_j^2 + y_j^2) 
 - \sum_{j=1}^{L-2} y_j x_{j+1}.
\end{equation}  
These states are $L$-fold degenerate, 
since `$e$', which represents the Cu with no hole, does not hop in this limit. 
The Schr\"odinger equation gives
\begin{equation}
  E x_j = E y_j = t_S ( x_{j+1} - 2 x_j - 2y_j + y_{j-1} )
\end{equation}
for $j=2,\cdots,L-2$ with boundary conditions 
\begin{eqnarray}
 E y_1 &=& t_S (x_2 - 2 x_1 -2 y_1), \nonumber\\
 E(y_1-x_1) &=& V x_1, \nonumber\\
 E y_{L-1} &=& t_S (x_{L-2} - 2 x_{L-1} -2 y_{L-1}), \nonumber\\
 E(y_{L-1}-x_{L-1}) &=& V x_{L-1}. \nonumber
\end{eqnarray}
The energy $E$ is measured from $\Delta$.
Due to symmetry $x_j$ and $y_j$ are real and there are even and odd parity 
solutions with $x_j=\mp y_{L-j}$. 

\begin{figure}
  \epsfxsize=15 truecm
  \centerline{\epsffile{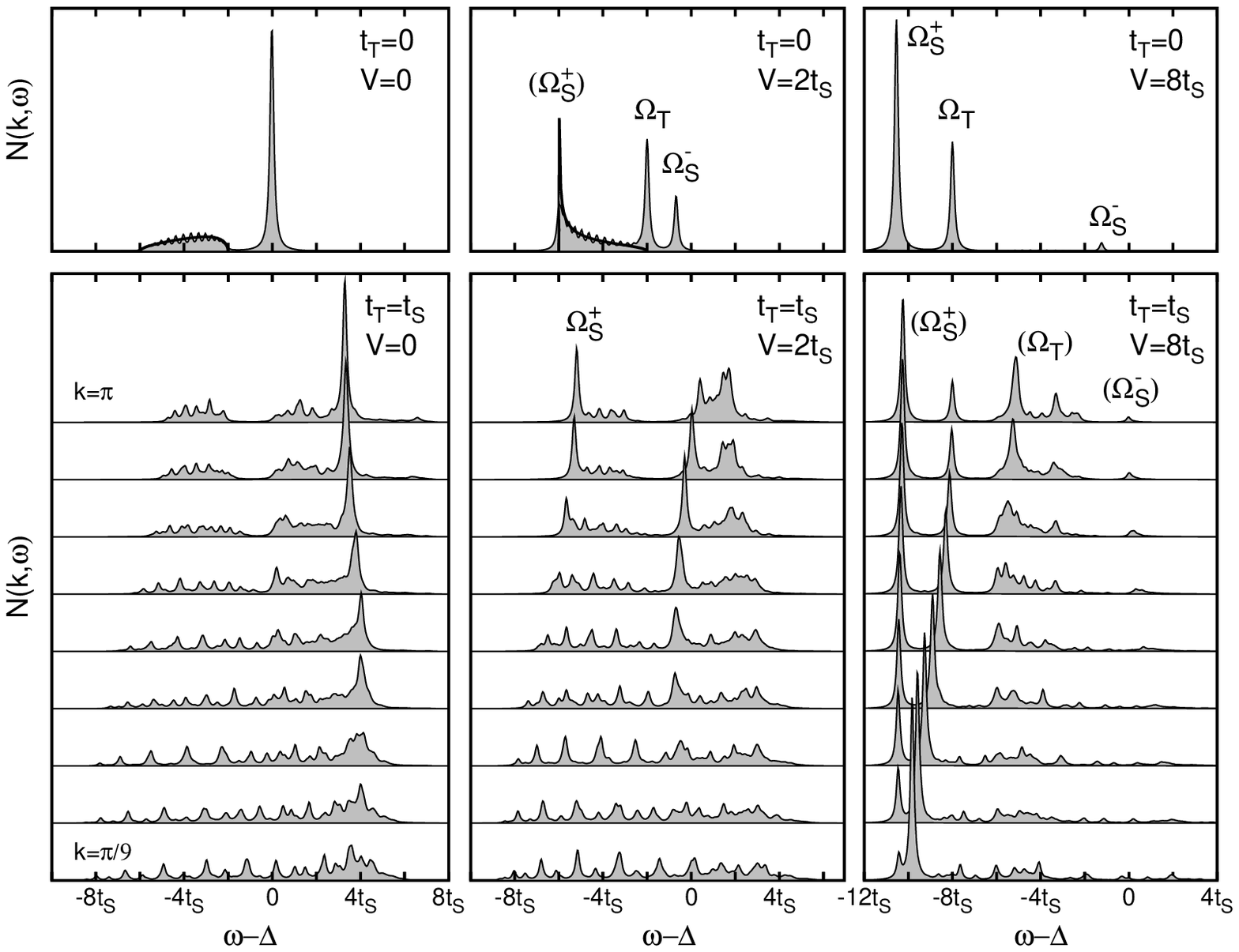}}
  \caption{
  ${\cal N}(k,\omega)$ of 18 site effective model for $V=0$, $2t_S$ and 
  $8t_S$ from left to the right, and $t_T=0$ (upper) and $t_T=t_S$ 
(lower plots) obtained by exact diagonalization. 
The thick line in the upper plots shows the analytical result.
  The $\delta$ functions are plotted as Lorentzians of width $0.1t_S$ and the 
plots for each $k$ are normalized to have total weight 1. In the upper plots
we show $k=\pi$ only because of the trivial $k$ dependence. The remnants 
of excitons are  indicated within the parenthesis in the lower right plot.
 These spectra are independent of the phase factors of Cu and O orbitals.
  }
  \label{fig:nok}
\end{figure}

 Let us first consider the case $V=0$. 
 We immediately see that $(x_j-y_j)E=0$, i.e. $x_j=y_j$ for
$E\neq 0$ and $j=1,\cdots,(L-1)$.
The solution is $x_j=y_j=\sin j \kappa$, with 
$\kappa = I \pi/L$, and  $I=1,2,\cdots,L-1$.
These states have even (odd) parity for $I$ even (odd),
energy 
\begin{equation}
  E_\kappa =  - 4 t_S + 2 t_S \cos \kappa,
\end{equation}
and norm $\langle f_S | f_S \rangle = L(2 - \cos \kappa)/2$. 
For $L\rightarrow \infty$ they will form a continuum 
from $E=-6t_S$ to $-2t_S$.
Additionally there are $L-1$ degenerate states with $E = 0$.
The matrix elements 
in Eq.~(\ref{eq:rhorho}) for the singlet contribution read:
\begin{equation}
 |\langle f_S | n_{k;\Delta} | {\rm GS} \rangle|^2 
 = \frac{8 t^2}{\Delta^2}  F_S 
      \frac{x_1^2 }{\langle f | f \rangle}
     \left[1 \pm \cos\frac{k}{2} \right]
      \sin^2 \frac{k}{4},
 \label{eq:mes}
\end{equation}
where the $+$ ($-$) sign is for the even (odd) state, and
$F_S=\langle {\rm GS} |  \frac{1}{4} - {\bf S}_0 \cdot {\bf S}_1 
 | {\rm GS} \rangle$ $\to \ln 2$ for $L \to \infty$ is the probability of
 finding two neighboring spins forming a singlet in the spin sequence. 
The only nontrivial quantity is 
$x_1^2/\langle f_S | f_S \rangle$, which can be conveniently expressed 
using the energy of the state as
\begin{equation}
  \frac{x_1^2}{\langle f | f \rangle} = 
  \frac{1}{L}\frac{2 \sin^2 \kappa}{2-\cos\kappa}
 =\frac{1}{L}
  \frac{
    \left(2t_S + E  \right) \left( 6t_S + E  \right)
  }{ t_S E }.
\end{equation}
In the thermodynamic limit we replace the sum over states 
with an integral over energy:
\begin{equation}
    \sum_I \rightarrow \int dE_\kappa \frac{1}{\pi}{\frac{\partial
    \kappa}{\partial E_\kappa}} ,
\end{equation}
where $(1/\pi)(\partial \kappa/\partial E_\kappa)$ is the density of states
\begin{equation}
 \frac{1}{\pi} \frac{\partial \kappa}{\partial E_\kappa} = 
 \frac{1}{\pi}\frac{1}{\sqrt{(-2t_S -E)( 6t_S + E )}}, 
 \label{eq:densta}
\end{equation}
the factor $\left[1 \pm \cos(k/2) \right]$ in Eq.~(\ref{eq:mes})
averages to 1, and for the contribution of the Zhang-Rice singlets to the 
density response we get:
\begin{equation}
 {\cal N}(k,\omega) =  \frac{8\ln 2}{\pi} \frac{t^2}{\Delta^2}
                       \frac{
                          \sqrt{
                            (\omega\!-\!\Delta\!+\!6t_S)
                            (\Delta\!-\!2t_S\!-\!\omega)
                          }
                       }{t_S (\Delta-\omega)}
                       \sin^2 \frac{k}{4}. 
\end{equation}
The density response in this limit has a trivial momentum dependence, 
due to the nondispersing nature of the `e' site.
It gives $2 (2-\sqrt{3}) F_S\approx  37\%$ of the total weight, 
the rest of the weight is in a single peak at $\omega=\Delta$, which comes
from the nondispersing singlets and triplets. The ${\cal N}(k,\omega)$ we just
calculated is shown on the upper left plot in Fig.~\ref{fig:nok}.

Turning on the Cu-O repulsion, which acts as an effective attraction between
the empty Cu site and the $O$ hole, two twofold degenerate 
(for $L \to \infty$) excitons with energies
\begin{equation}
   \Omega^{\pm}_S = \frac{3 t_S V+V^2 \pm V \sqrt{12 t_S^2+V^2} }{t_S-2V}
\end{equation}
appear, together with a twofold degenerate exciton involving the triplets at 
$\Omega_T=-V$. 
The $\Omega^{+}_S$ solution exists only for $V>2t_S$ where it 
splits off from the lower edge of the continuum. Not going into the details, 
the expression for $x_1^2/\langle f_S | f_S \rangle$ in Eq.~(\ref{eq:mes}) 
now reads
\beginwideequ
\begin{equation}
  \frac{x_1^2}{\langle f | f \rangle} = \frac{
  -E \left(2t_S + E  \right) \left( 6t_S + E  \right)
}{
  24 t_S^2 V +  2E(4t_S -  V)V + 
  L \left[ (2 V - t_S) E^2 + 2 (3 t_S + V)V E + 3 V^2 t_S
      \right]},
 \label{eq:x1V}
\end{equation}
\endwideequ
which is valid both for the $\Omega^{\pm}_S$ excitons and the continuum. The 
${\cal N}(k,\omega)$ is complicated and we do not give the analytical form, 
which is straightforward to derive from Eqs.~(\ref{eq:mes}), (\ref{eq:densta}),
and (\ref{eq:x1V}), but we refer to 
Fig.~\ref{fig:nok} for a discussion of features.
For small $V$ the energy of the 
exciton is $\Omega^-_S \approx (3-2 \sqrt{3}) V$ with 
relative weight $\approx\left[ 2 \sqrt{3}-3 - 2 (7-4\sqrt{3}) V/t_S \right] 
 F_S$, i.e. increasing $V$ the weight is transferred 
to the continuum. For large repulsion ($t_S\ll V\ll U$) 
the weight is concentrated in the 
excitons at $\Omega^+_S \approx -V-2t_S$, while the continuum and the 
exciton at $\Omega^-_S \approx -3t_S/2$ has a negligible weight of 
the order of $t_S^2/V^2$. 
On the other hand, the triplet exciton $\Omega_{T}$ has 
weight $1-F_S\approx 31\%$, independent of the size of $V$.

 To study the effect of finite $t_T$, we used numerical L\'anczos diagonalization of small systems (18 site) to extract the density-density correlations. 
As can be seen in Fig.~\ref{fig:nok}, the effect of finite $t_T$ is 
substantial: 
(i) Because `$e$' acquires dispersion, the shape 
of the spectrum resembles more closely that of the one--band model, in that it 
narrows as $k \to \pi$\cite{hubbard}. Turning on $V$, 
excitonic features are formed at the zone boundary. 
(ii) The contributions coming from `singlets' and `triplets' cannot be
separated energetically, especially for small momenta. 
(iii) While for $t_T=0$ 
the excitons are sharp peaks, for $t_T \neq 0$ these sharp excitonic peaks 
broaden and form an incoherent spectrum.
A direct comparison with the experiments to decide whether a two band model 
is more appropriate is difficult, given that the resolution of the EELS 
spectra is not sufficient to see the detailed differences between the two band 
and the one band models.  Furthermore, all of our results here are valid in
the strong-coupling limit, so one expects quantitative changes to the spectra
for experimentally relevant coupling strengths.

\end{multicols}
\widetext
\begin{multicols}{2}
\narrowtext

 Recently, the density response was calculated using projection techniques in
 Ref.~\onlinecite{richter}, with a result which disagrees with ours.
 Since the calculation presented in that paper is rather involved, it is 
difficult to trace whether the difference is due to the method applied or 
to the parameter regime investigated.

\section{conclusion}

 We have demonstrated that spin-charge factorization 
 may be applied to understand dynamical behavior of the two-band model in a 
particular limit. The low energy hole charge carriers have been identified 
as complex objects 
resembling Zhang-Rice singlets, and the low energy part of the 
single-particle spectral function of the two-band model has been shown to 
be related to that of the one-band model with nontrivial frequency as well as 
momentum dependent corrections. 
This provides a very simple and natural explanation for the momentum and
frequency dependence of the spectral weights observed experimentally.

\begin{acknowledgments}

We would like to thank Prof. P. Fulde for his kind hospitality in 
Max-Planck-Institute f\"ur Physik komplexer Systeme in Dresden, where 
the present work started.
This work was part-funded by Hungarian OTKA D32689, AKP98-66, 
and Bolyai 118/99.
\end{acknowledgments}

\end{multicols}

\end{document}